\documentclass[]{aa}  
\usepackage{subfigure}

\usepackage{graphicx}
\usepackage{txfonts}
\usepackage{xcolor}
\usepackage{natbib,twoopt}
\usepackage[breaklinks=true]{hyperref} %% to avoid \citeads line fills
\bibpunct{(}{)}{;}{a}{}{,}             %% natbib format for A&A and ApJ
\makeatletter
  \newcommandtwoopt{\citeads}[3][][]{\href{http://adsabs.harvard.edu/abs/#3}%
    {\def\hyper@linkstart##1##2{}%
     \let\hyper@linkend\@empty\citealp[#1][#2]{#3}}}
  \newcommandtwoopt{\citepads}[3][][]{\href{http://adsabs.harvard.edu/abs/#3}%
    {\def\hyper@linkstart##1##2{}%
     \let\hyper@linkend\@empty\citep[#1][#2]{#3}}}
  \newcommandtwoopt{\citetads}[3][][]{\href{http://adsabs.harvard.edu/abs/#3}%
    {\def\hyper@linkstart##1##2{}%
     \let\hyper@linkend\@empty\citet[#1][#2]{#3}}}
  \newcommandtwoopt{\citeyearads}[3][][]%
    {\href{http://adsabs.harvard.edu/abs/#3}
    {\def\hyper@linkstart##1##2{}%
     \let\hyper@linkend\@empty\citeyear[#1][#2]{#3}}}
\makeatother
\newcommand{\orcid}[1]{}
\begin{document}

   \title{Inferring intrahalo light from stellar kinematics}
   \subtitle{A deep learning approach}

   \author{I.~Marini
          \inst{1}
          \and
            A.~Saro\orcid{0000-0002-9288-862X}\inst{2,3,4,5,6}
          \and
          S.~Borgani\orcid{0000-0001-6151-6439}\inst{2,4,3,5,6}
          \and
          M. Boi\inst{2}
          }

   \institute{European Southern Observatory, Karl Schwarzschildstrasse 2, 85748, Garching bei M\"unchen, Germany\\
                \email{ilaria.marini@eso.org}
         \and
            Dipartimento di Fisica - Sezione di Astronomia, Universit\`a di Trieste, Via Tiepolo 11, 34131 Trieste, Italy
        \and 
            INAF-Osservatorio Astronomico di Trieste, Via G. B. Tiepolo 11, 34143 Trieste, Italy
        \and
            IFPU, Institute for Fundamental Physics of the Universe, Via Beirut 2, 34151 Trieste, Italy
        \and
            INFN, Sezione di Trieste, Via Valerio 2, 34127 Trieste, Italy
        \and
            ICSC - Italian Research Center on High-Performance Computing, Big Data and Quantum Computing
        \and 
             Data Science and Scientific Computing, Università di Trieste,
             Via A. Valerio 12/1, 34127 Trieste, Italy
             }

   \date{Received  16 February 2024 / Accepted 28 June 2024}

% \abstract{}{}{}{}{} 
% 5 {} token are mandatory
 
  \abstract
   {In the context of structure formation, disentangling the central galaxy stellar population from the stellar intrahalo light can help us shed light on the formation history of the halo as a whole, as the properties of the stellar components are expected to retain traces of the formation history. Many approaches are adopted to assess the task, depending on different physical assumptions (e.g. the light profile, chemical composition, and kinematical differences) and depending on whether the full six-dimensional phase-space information is known (much like in simulations) or whether one analyses projected quantities (i.e. observations). }
   {This paper paves the way for a new approach to bridge the gap between observational and simulation methods. We propose the use of projected kinematical information from stars in simulations in combination with deep learning to create a robust method for identifying intrahalo light in observational data to enhance understanding and consistency in studying the process of galaxy formation.}
   {Using deep learning techniques, particularly a convolutional neural network called U-Net, we developed a methodology for predicting these contributions in simulated galaxy cluster images. We created a sample of mock images from hydrodynamical simulations (including masking of the interlopers) to train, validate and test the network. Reinforced training (Attention U-Net) was used to improve the first results, as the innermost central regions of the mock images consistently overestimate the stellar intrahalo contribution. }
   {Our work shows that adequate training over a representative sample of mock images can lead to good predictions of the intrahalo light distribution. The model is mildly dependent on the training size and its predictions are less accurate when applied to mock images from different simulations. However, the main features (spatial scales and gradients of the stellar fractions) are recovered for all tests. While the method presented here should be considered as a proof of concept, future work (e.g. generating more realistic mock observations) is required to enable the application of the proposed model to observational data. }
   {}

   \keywords{Methods: data analysis --
                Techniques: miscellaneous --
                Galaxies: stellar content
               }

   \maketitle
%
%-------------------------------------------------------------------
\section{Introduction}
According to the standard cosmological model, galaxy clusters and groups build up their stellar mass through star formation and subsequent mergers \citep{pillepich_first_2018,ragone-figueroa_bcg_2018,montenegro-taborda_growth_2023}. This two-phase scenario also naturally explains the relative old stellar population and passive star formation in central galaxies, where clusters still accrete at $z=0$. Because of this mechanism, most of the stars are locked up in the central (and satellite) galaxies \citep{kravtsov_formation_2012} orbiting in clusters and groups, while a considerable fraction is composed of free-floating stars bound to the halo potential which, for its collisionless nature \citep{binney_galactic_2011}, is an exemplary fossil of the cluster formation history. The intrahalo light, whose dependence on the mass will determine whether it is defined intracluster light (ICL\footnote{for simplicity, we refer to both group and cluster-size halo as more generally ICL}) or intragroup light \citep{montes_intracluster_2019,alonsoasensio_intracluster_2020,montes_faint_2022, arnaboldi_kinematics_2022}. A typical example is the imprints left in the outer stellar region of unique morphological features, such as shells, ripples, and tidal tails \citep{bilek_census_2020, montes_new_2022, valenzuela_stream_2022} as a natural result of merger events and/or radial infall of satellites \citep{pop_formation_2018,karademir_outer_2019}.
\par
One of the most prominent unsolved issues in this field is how to effectively separate the light from the Brightest Cluster Galaxy (BCG\footnote{We name the central stellar component BCG regardless of the size of the host halo which could be in the cluster or group regime interchangeably.}) and ICL since they share similar spatial scales. In recent years, several studies \citep[e.g.][and references therein]{contini_transition_2022,montes_faint_2022} have discussed the role of the transition radius, the distance at which the ICL component starts to dominate the stellar component. Due to the variety of methods employed to estimate the ICL contribution, the value of this transition radius may depend on the adopted method of ICL identification. From the observational side, typical values of the transition radius are around 60 $-$ 80 kpc \citep{montes_buildup_2021,gonzalez_discovery_2021}, thus in line with results from earlier works \citep[e.g][]{zibetti_intergalactic_2005,gonzalez_census_2007}. These values slightly increase for other analyses, such as those presented by \cite{zhang_dark_2019}, who concluded that the transition from the BCG to the ICL is just outside 100 kpc, or by \cite{chen_sphere_2022} who found values ranging in the interval 70 $-$ 200 kpc. Results based on semi-analytical simulations \citep[e.g.][]{contini_brightest_2021,contini_origin_2021,contini_transition_2022} agree with these observational results, and indicate that the transition radius is independent of both BCG$+$ICL and halo masses, with typical values of 60 $\pm$ 40 kpc, if similarly derived from profile fitting. Usually, this technique requires the assumption of a double or triple S\'ersic profile \citep{sersic_influence_1963} or a composition of different profiles such as the Jaffe profile \citep[][describing the BCG distribution;]{jaffe_simple_1983} and NFW profile \citep{navarro_universal_1997} for the ICL \citep{kluge_photometric_2021}. 
\par
In hydrodynamical simulations, this task is eased thanks to the possibility of accessing the full six-dimensional phase-space traced by the stellar particles. \cite{proctor_identifying_2023} used Gaussian mixture methods to decompose the stellar halo into three components (i.e. a disc, a bulge, and ICL) according to their kinematic properties. The authors find the observational equivalent of the transition radius at approximately 30 kpc for halos with $\log (M_{200}/ M_{\odot})<12.8$ which quickly increases for higher masses. Likewise, \cite{dolag_substructures_2009} and  \cite{marini_machine_2022} have demonstrated how an unbinding procedure can be applied to the stellar halos of galaxy clusters to yield two separate kinematical subsets traceable to the BCG and the ICL. The assumptions underlying this method are derived from the velocity distribution of star particles that exhibit a bimodal distribution associated with two dynamically distinct components. Combining this information with an unbinding procedure leads to separation into a central BCG (more compact and dynamically cold) and a hotter diffuse ICL. This kinematical distinction is hardly transferable to observations as shown in \cite{remus_outer_2017} since the kinematical distinction does not necessarily imply a dual and clear distinction of the radial surface density profile.
\par
Thus, it is undeniable that there is a substantial gap between the identification methods applied in simulations (fundamentally based on the complete knowledge of the kinematical and positional information of the stellar component) and in observations \citep{rudick_quantity_2011, kluge_photometric_2021, arnaboldi_dynamics_2001}, further compromising the efforts to uniform studies on the topic \citep{montes_faint_2022}. In the local Universe, a viable path is to connect high-resolution integral field spectrography (IFS) observations (to extract the fine-grained kinematical structure) with the outcome of numerical simulations. This work aims to pave the way for this perspective using deep learning (DL) techniques \citep[][for a review]{lecun_deep_2015}. Since its breakthrough \citep{krizhevsky_imagenet_2012}, DL has rapidly gained acclaim in many scientific applications, including astronomy \citep{carleo_machine_2019, smith_astronomia_2023} since the usual data volume naturally favours the applications of these techniques. Many applications include improvements in the estimates of photometric redshifts \citep[e.g.][]{collister_annz_2004, feldmann_zurich_2006, salvato_dissecting_2011}, galaxy morphology identification \citep{dieleman_rotation-invariant_2015, ball_galaxy_2004, banerji_galaxy_2010}, exoplanet detection \citep{gibson_gaussian_2012}, gravitational wave physics \citep{george_deep_2018}, and analysis of the stellar galactic disc \citep{cantat-gaudin_painting_2020}, just to name a few. In recent years, a breakthrough in cosmological studies has been reached with the advent of machine learning-based cosmological simulations \citep{he_deep_2015, kamdar_machine_2016, villaescusa-navarro_quijote_2020} further proving the advancements in the field. In this complex scenario, convolutional neural networks \citep[CNNs;][]{lecun_gradient-based_1998, krizhevsky_imagenet_2012} have played an instrumental role in astronomy for their ability to automatically learn spatial features from raw pixel data, making them highly effective in tasks such as image classification \citep{dominguez_sanchez_improving_2018}, object detection \citep{schanche_machine-learning_2019}, and image segmentation \citep{burke_deblending_2019}. 
\par
In this paper, we illustrate a new method devoted to recovering the projected ICL distribution from images of simulated galaxy clusters using CNNs. By leveraging DL techniques, we can accurately extract information about the ICL from images, facilitating our understanding of its properties and distribution within galaxy clusters. We point out that the main purpose of this work is to present a proof of concept to identify ICL by exploiting information on the stellar kinematics, rather than delivering a stand-alone method ready to be applied to observational data. For this reason, several limitations inherent in the approach presented here (e.g. the lack of a fully realistic observational image) will need to be addressed with further investigation before the proposed method can be effectively applied to observational data. On the other hand, it is important to note that observers will increasingly require improved methods for detecting and characterising the ICL in real observational data. As observational capabilities continue to evolve, the development of advanced CNN-based methods will be essential for unlocking the full potential of ICL studies in observational astronomy. Furthermore, the efficacy of this method will rely on its adaptation to more realistic simulations of the stellar structure within galaxy clusters and mock observations. Incorporating detailed models that account for various physical processes affecting stellar populations and the observational process itself will enhance the fidelity of the CNN-based approach, ensuring its applicability to a wide range of observational scenarios and contributing to a deeper understanding of the ICL in the context of galaxy cluster evolution.
\par
The paper is structured as follows. In Sect.~\ref{sec:methods}, we describe the main ingredients of our modelling: the simulation set, the mock images extracted, and the DL models. In Sect.~\ref{sec:results}, we present the main results of the study, caveats, and future perspectives; in Sect.~\ref{sec:conclusions} we summarise our findings and draw the main conclusions of our analysis. 

\section{Methods}
\label{sec:methods}

\subsection{Input simulations}
In this section, we describe the suite of simulations (i.e. Dianoga) used to construct the mock catalogue. Dianoga is a set of zoom-in cosmological hydrodynamical simulations of galaxy clusters carried out with GADGET-3, a Tree Particle Mesh -- smoothed particle hydrodynamics (SPH) code, which represents the evolution of the public GADGET-2 code \citep{springel_cosmological_2005}. The most important changes in our developer branch of GADGET-3 include the use of a higher-order kernel interpolating function, time-dependent artificial viscosity, and artificial conduction schemes, which in turn alleviate several limitations of standard SPH implementations \citep{dolag_turbulent_2005,beck_improved_2016}. 
\par
\begin{figure}
    \centering
    \includegraphics[scale=0.5]{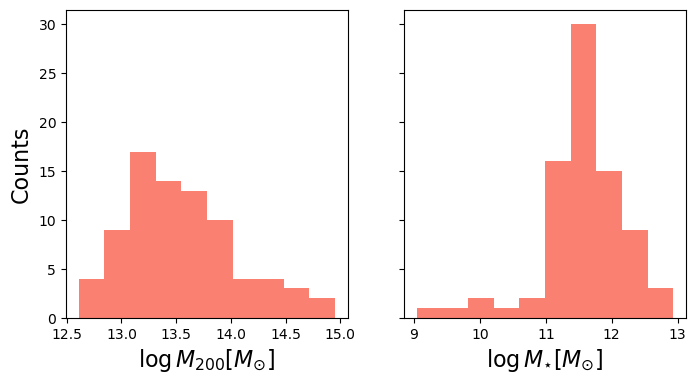}
    \caption{Halo mass (left) and stellar mass (right) distribution of the 80 halos selected within the eight Dianoga Lagrangian regions at $z=0.3$. }
    \label{fig:Mass_Distributions}
\end{figure}

At the resolution of choice, the set comprises a total of eight Lagrangian regions which have been selected around some of the most massive halos in a lower-resolution N-body parent box of comoving side of 1 $h^{-1}$ cGpc. Each region hosts several halos in both groups and clusters-mass regimes: for this project, we select each such region among the ten most massive halos. They sum to a total of 80 halos within the group and cluster regime (i.e. $\log M_{200}\sim 10^{13}-10^{14} M_{\odot}$). In Fig.~\ref{fig:Mass_Distributions}, we plot the halo mass (left panel) and stellar mass distribution (right) for these halos. Their physical properties are extensively discussed in \cite{bassini_dianoga_2020} and in \cite{marini_velocity_2021}.
\par
Initial conditions have been generated following the prescription in \cite{tormen_adding_1996} for a $\Lambda$ Cold Dark Matter ($\Lambda$CDM) cosmological model with $\Omega_\mathrm{M}=0.24$, $\Omega_\mathrm{b}=0.04$, $n_s=0.96$, $\sigma_8 = 0.8$ and $H_0=72$ km s${^{-1}}$ Mpc${^{-1}}$. In the highest-resolved regions, the DM particle mass is $m_{DM} = 8.3\times10^7 \, h{^{-1}}$ M$_{\odot}$ and the initial gas particle is $m_{gas} = 3.3\times10^7 \, h{^{-1}}$ M$_{\odot}$. The Plummer equivalent length for the DM particles corresponds to $ \epsilon=3.75\, h{^{-1}}$ kpc, whereas gas, stars and black hole particles have $\epsilon=3.75\, h{^{-1}}$ kpc, $1 \, h{^{-1}}$ kpc and $1\, h{^{-1}}$ kpc at $z=0$, respectively. 
\par
Several subgrid models describe the unresolved baryonic physics of the simulations, including radiative cooling, star formation, and stellar feedback \citep{springel_cosmological_2003}, metal and chemical enrichment \citep{tornatore_chemical_2007}, and Active Galactic Nuclei (AGN) feedback \citep[for more details, see Appendix A in][]{ragone-figueroa_brightest_2013}. Further details on the simulation can be found in \cite{bassini_dianoga_2020}. 

\subsubsection{Identification of structures}
The catalogue of particles associated with structures (e.g. halos of groups and clusters of galaxies) and substructures (i.e. subhalos or galaxies) in simulations is compiled by a halo-finder. Firstly, a friend-of-friend (FoF) algorithm is run on the particles: this procedure records an initial guess on the hierarchical structure of the simulation based on geometrical assumptions. If particles are clustered in samples with their inter-particle distance smaller than the linking length ($b=0.2$ in units of mean inter-particle separation, in our case) their unique identification number is stored in association with a halo. A fundamental drawback of this method is that it will occasionally link independent structures together across particle bridges. Furthermore, it will only identify large systems, leaving smaller structures (i.e. subhalos) in dense environments unrecorded. Therefore it is instrumental to benchmark the potential subhalo catalogue through an unbinding procedure, such as the one provided by {\tt SubFind} \citep[]{springel_populating_2001,dolag_substructures_2009}. The algorithm runs on the single FoF-identified halos basing its decision on an excursion set procedure. By descending along the density gradient, {\tt SubFind} creates a list of potential subhalo candidates whose binding energy is later investigated. This amounts to eliminating those particles whose energy makes them unbound to the substructure: if more than a certain minimum number of particles (50) survive the unbinding procedure, the substructure is identified as a genuine subhalo. The centre of each subhalo is identified with the position of the member particles having the minimum value of the gravitational potential. The properties of the halos and subhalos are then determined based on the properties of the particles composing them. 

\subsubsection{Identification of the ICL}
\label{subsec:icl_identification}
A second unbinding procedure can be applied to the stellar particles bound to the main halo, as they represent the contribution from both the BCG and the ICL. Theoretical predictions and observational results (see \citealt{dolag_dynamical_2010})  have shown that the two components populate different regions of the phase-space. The classification is achieved on each single star particle through the automatic classification performed by a random forest \citep{marini_machine_2022} trained on classified particles by {\tt ICL-SubFind} \citep{dolag_dynamical_2010}. The original algorithm assumes the double Maxwellian found in the three-dimensional particle velocity distribution of the two stellar components to be two single Maxwellian distributions associated with the ICL and the BCG, respectively \citep{murante_diffuse_2004}. More specifically, the ICL is associated with the Maxwellian yielding the largest velocity dispersion, in contrast, the BCG, having colder dynamics, populates the distribution at lower dispersion. To assign each star particle to either of the two dynamical components, the algorithm follows an unbinding procedure comparing its kinetic and potential energy to the assumed potential energy of the central subhalo. Reasons to prefer the random forest to the original algorithm for the separation between ICL and BCG components from its stability to its faster response. More details can be found in \cite{marini_machine_2022}.

\subsection{Generation of the input images}
\label{subsec:mock_images}
\begin{figure}
    \centering
    \includegraphics[scale=0.5]{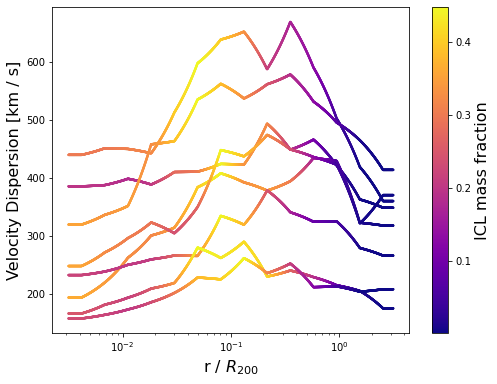}
    \caption{Velocity dispersion profiles of the most massive halos in each of the eight Lagrangian regions. The profiles are displayed as a function of the radial distance from the centre normalised for $R_{200}$ of the halo. The profiles are colour-coded according to the mean ICL mass fraction within the spherical annuli.}
    \label{fig:10percR200}
\end{figure}

The design of the images based on our set of simulated clusters mimics the geometrical conditions of a hypothetical observation by a state-of-the-art IFS. On one hand, the difficulties of observing the light at low surface brightness composing the ICL naturally favours close targets, and thus zoomed images. On the other hand, we need to ensure that the hypothetical field of view (FOV) of the IFS is large enough to include the physical scales at which our simulations predict a significant gradient in the ICL presence with increasing radius. To this purpose, in Fig.~\ref{fig:10percR200} we show the distribution of the velocity dispersion profiles (along the line of sight) of the most massive halos in each of the eight Lagrangian regions and we estimate the transition radius defined as the physical radius at which the ICL fraction dominates over the BCG+ICL stellar mass.
We notice that \cite{marini_velocity_2021} have shown that this set of groups and clusters has velocity dispersion profiles that agree with highly resolved spectroscopic observations of nearby clusters. To measure these profiles, we use all the stars bound to the halo within a cylinder of length $2R_{200}$\footnote{We define $R_{\Delta}$ as the radius encompassing a mean halo overdensity equal to $\Delta$ times the critical density of the universe at a given redshift $\rho_c(z)$.} around the centre of the halo. This leads to including stars bound to subhalos (i.e. satellite galaxies), but not strictly belonging to BCG or ICL, which results in conditions close to the observational ones. Curves are colour-coded according to the ICL mass fraction (defined as the ICL stellar mass in a radial bin over the total stellar mass of the bin) as a function of the clustercentric distance in $R_{200}$ units. The vertical scatter in the velocity dispersion profiles is explained by the different halo masses. We observe the consistency in the ICL mass fraction with radial distance as in most of the profiles this fraction is peaking at around $0.1\, R_{200}$, corresponding to the transition radius. This fraction decreases for larger radii as the stellar component associated with the subhalos starts dominating. This suggests that observations should cover a large enough area to include this radius to detect the transition from the BCG to the ICL regimes. Furthermore, this peak is often consistent with the peak in the velocity dispersion profiles, whereas for larger radii the fraction is reduced due to the presence of substructures. Since the typical value of this transition radius is nearly constant once expressed in terms of $R_{200}$, it implies that this quantity is mainly connected to the cosmological build-up of the halo. This result is consistent with the findings in \cite{proctor_identifying_2023} on a C-EAGLE sample of groups and clusters of galaxies of similar halo masses. 
\par
 
To set the spatial resolution and FOV within which we will be conducting our study, it is required to finalise the choice of a specific detector. We choose to simulate the geometrical conditions of the instrument MUSE\footnote{MUSE official webpage \url{https://www.eso.org/sci/facilities/paranal/instruments/muse.html\#par_title}} \citep{bacon_muse_2010} at the VLT which can guarantee the capabilities of Integral Field Unit (IFU) in a wide FOV: the IFU has a FOV of 1 arcmin$^2$ and resolution of $(0.2\times0.2)$ arcsec$^2$. This choice is primarily driven by constructing a realistic geometrical setup and does not imply that other observational conditions (e.g. the spectral features) are used in this work. This implies that even though throughout the paper we will speak of mock images, we do not claim to reproduce the necessary observational conditions to be as such. In physical units, this setup translates into $(263.34 \times 263.34)$ kpc$^2$ and resolution $(2\times2)$ kpc$^2$ at $z=0.3$, for our reference cosmological model. In order to be conservative on the resolution effect, we double the image size and pixel resolution to $(526.68 \times 526.68)$ kpc$^2$ and $(4\times4)$ kpc$^2$. In this setting, each image has $130\times130$ pixels. For the only purpose of speeding up the training of our DL algorithm, we crop the images to the nearest power of 2 in pixels (i.e. $128\times128$), following the indications in \cite{goodfellow_deep_2016}. We choose this frame size for all objects thus covering different dynamical ranges for clusters and groups having different masses: we reach $R_{200}$ for the smallest group with $\log(M_{200}/M_{\odot}) = 12.75$, and for the largest cluster ($\log(M_{200}/M_{\odot}) = 15.10$) we cover only the central $0.1\, R_{200}$.
\par
We create the images to be the post-processed line-of-sight velocity dispersion maps of the star particles in each system, much like what is extracted from the broadening of the spectral lines in the IFU images. To guarantee consistency with the kinematical structure of each system, we reconstruct the velocity dispersion map of all the stars in the frame. Then we mask the pixels containing contaminating galaxies from our images thanks to the halo structure provided by {\tt SubFind}. In order words, we simulate the standard procedure to mask foreground and background galaxies in data analysis. We follow the same procedure to create the ICL mass fraction image (i.e. the ground truth) on which we will train the network. 
\par
One of the caveats connected with DL techniques is the necessity of a large sample size to effectively train and test a model. Choosing only one projection from the largest halos in each Lagrangian region (amounting to eight images) would not suffice the demand, therefore, we are bound to actively expand our dataset with data augmentation. We start by investigating the ten largest halos in our sample for each Lagrangian box (i.e. 80 halos). Then, we take advantage of the halo triaxiality and select for each one 26 different line-of-sight projections which in the three-dimensional space corresponds to taking 45 degrees rotation from the initial system of reference. These choices already guarantee us 2080 images however we further extend it by asking that our images may not be centred on the cluster centre but slightly shifted by a random offset within $15\%$ of the image size. This operation is performed three times. We further discuss the role of the data size in Sect.~\ref{subsec:size of the training set}.
\par
The final dataset comprises 6240 synthetic images that are split into $70:20:10$ subsets to create training:validation:testing sets. Additionally, we create a smaller testing set extracted from ten halos in a different Lagrangian region not previously used for the experiment. The advantage is that this testing set is independent of the halos used to train and validate the DL algorithm. We perform only two rotations around the main axes and two centre-shifts for a total of 40 images. We will present here only the results for this smaller testing set, since it proves to be consistent with the larger more-dependent testing set.  
\par
\begin{figure*}
    \centering
    \includegraphics[width=\linewidth]{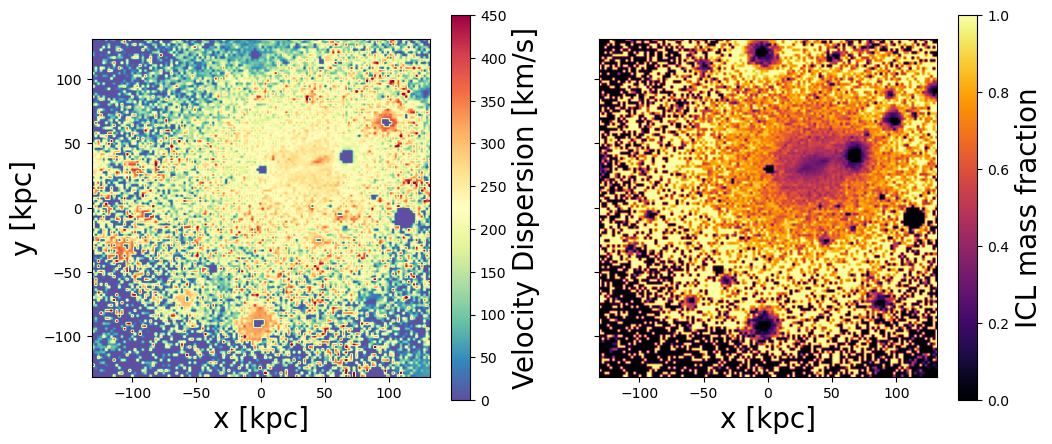}
    \caption{Examples of the mock images. In the left panel, we show a velocity dispersion map, and in the right panel the corresponding maps of ICL mass fraction from a randomly extracted halo.}
    \label{fig:example}
\end{figure*}

An example of images for the velocity dispersion (left panel) and the ICL mass fraction (right panel) is presented in Fig.~\ref{fig:example}. There is a mild correlation between the ICL mass fraction and the ICL velocity dispersion map which we exploit for this experiment. This can be hardly appreciated with the naked eye (some features are visible around the BCG position and going outwards) but we already discussed the role of the kinematical profiles of the BCG and ICL. Therefore, we anticipate that a CNN will be able to detect these signals in the simulated images. 

\subsection{The U-Net architecture}
In our work, we aim to predict the ICL fraction in mock images of galaxy clusters by exploiting the information on the velocity dispersion of the stellar component, as provided by an IFS. In more general terms, the desired output is an image (or multidimensional object) rather than a single class label. 
\par
Many examples from this class of problems are present in biomedical applications, as image diagnosis is sensitive to a variety of scales in health-related problems. In many of these studies \citep[for a review]{siddique_recurrent_2021} the preferred DL architecture has been the U-Net \citep{ronneberger_u-net_2015}, while later it further gained importance in other fields \citep[e.g. in astronomy][]{vojtekova_learning_2021, chadayammuri_painting_2023}. The architecture consists of two main branches: a contracting path said to capture the general context and a symmetric expanding path that enables precise localisation (see Fig.~1 in \citealt{oktay_attention_2018} for a schematic representation of the architecture). This double action increases the resolution of the output: high-resolution features from the contracting path are combined with the upsampled output allowing better localisation of the features. The two branches are connected through several skip connections \citep[][the first use of skip connection is attested in resnet;]{he_deep_2015} which correspond to bypassing some of the neural network layers and feed the output of one layer as the input to the following levels. It is a standard module and provides an alternative path for the gradient with backpropagation. The idea is to skip connections in different points of the architecture to allow fine-grain recollection of the original image, throughout the learning process. The contracting path is built following the typical architecture of a CNN: a sequence of $3\times 3$ convolution matrices \citep[with ReLU activation functions\footnote{A ReLU (rectified linear unit) activation function is an activation function that introduces the property of non-linearity to a deep learning model. Its mathematical expression is $ReLU(z) = max(0,z)$};][]{glorot_deep_2011,agarap_deep_2019} and $2\times 2$ max pooling operations to downsample. After each max pool, the feature channels are doubled (as shown at the top of the rectangular boxes). Every step in the expansive path consists of an upsampling of the feature map followed by a $2\times 2$ convolution that halves the number of feature channels and two $3\times 3$ transposed convolutions, each followed by a ReLU. In the final layer, a $1\times 1$ convolution is used to map each $64$-component feature vector to the original $128\times128$ image. The output is passed through a sigmoid function\footnote{A sigmoid} function is a mathematical function having a characteristic S-shaped curve or sigmoid curve. Its mathematical expression is $\sigma(z) = \exp^z/ (1+\exp^z)$, to map each pixel between 0 and 1.
\par
The second network we used in this work is the Attention U-Net \citep{oktay_attention_2018,schlemper_attention_2019} an evolution of the U-Net architecture since it integrates attention mechanisms to enhance its ability to capture relevant features and improve the quality of image segmentation. The attention gates \citep{jetley_learn_2018} induce the network to focus on the regions of interest with no additional supervision nor significant computational overhead.
By incorporating attention mechanisms within the skip connections between the contracting and expansive paths of the U-Net, the relevant information is filtered, enabling the translation of intricate patterns and textures while suppressing background noise. The attention mechanism acts as a dynamic filter, adaptively modulating the importance of different spatial locations across feature maps. In this work, we will use both Attention U-Net and the original U-Net.
\par
Our model is written with the support of PyTorch \citep{paszke_pytorch_2019}. Unless stated otherwise, the model's parameters (i.e. weights and biases) are initialised according to the default values of the library. The loss function is the mean squared error (MSE) between the true label $y_i$ and the predicted one $\hat{y}_i$ which is formally defined by the following equation:
\begin{equation}
    \mathrm{MSE} = \frac{1}{N} \sum_{i=1}^{N} (y_i -\hat{y}_i)^2,
\end{equation}where $N$ is the number of samples we are testing against.
Initially, the model was compiled with the Adam optimiser \citep{kingma_adam_2017} following the pre-training learning rate range test described by \cite{smith_cyclical_2017} to find an optimal static learning rate. In short, this method prescribes an initial phase of parameter-tuning where we vary the learning rate at each epoch for a value that is exponentially (or linearly) increased within a given interval. Changing the learning rate, and consequently studying the change in the loss curve, can help in understanding the optimal value to plug in during the effective training phase. The optimal value is defined as the point at which the loss function is the steepest since the learning will be more efficient at this point.
\par
Later, we found the One-Cycle policy \citep{smith_super-convergence_2018} with the stochastic gradient descent (SGD) to give even more satisfactory results, so we settled for this method. This approach anneals the learning rate from an initial value to some maximum learning rate and then back to some minimum learning rate much lower than the initial learning rate. Thus, the main difference is the dynamic learning rate.

\section{Results}
\label{sec:results}
\begin{figure*}
    \centering
    \includegraphics[scale=0.68]{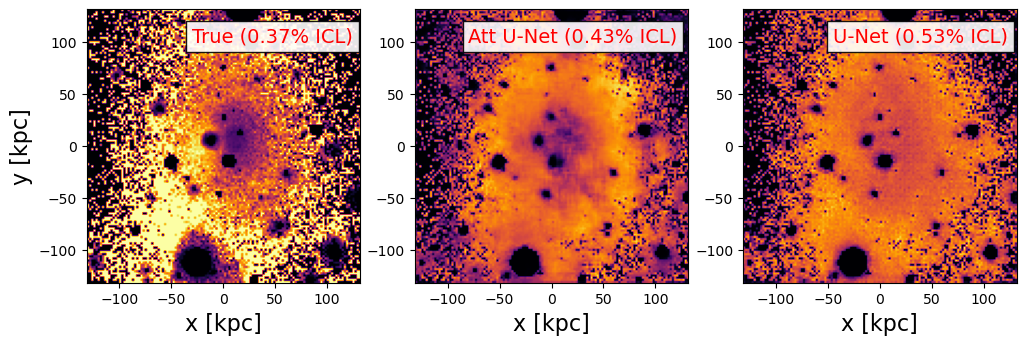}
\end{figure*}
\begin{figure*}
    \centering
    \includegraphics[scale=0.68]{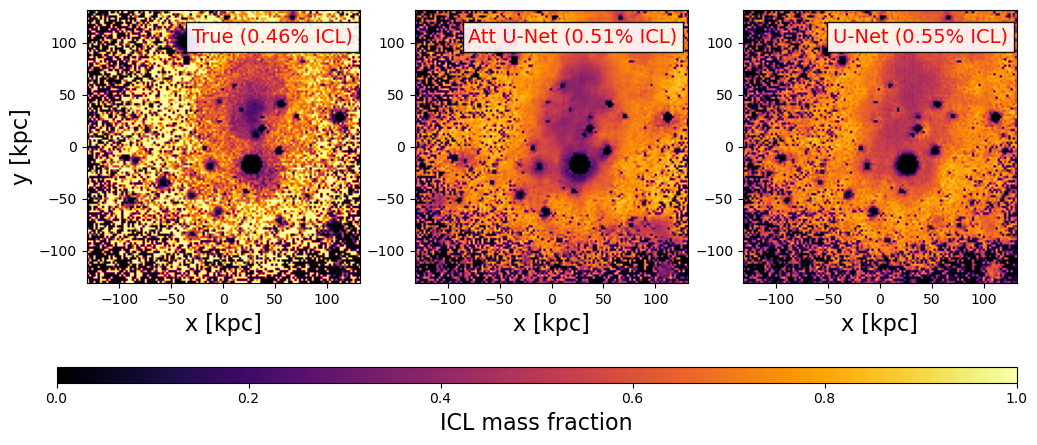}
    \caption{ICL mass fraction extracted with {\tt ICL-SubFind} (left panels), the Attention U-Net (central) and U-Net (right) from two random halos extracted from the test set. The colour map highlights the fraction within each image. We also report within parenthesis the ICL mass fraction within each image. }
    \label{fig:result3}
\end{figure*}
\begin{figure}
    \centering
    \includegraphics[scale=0.68]{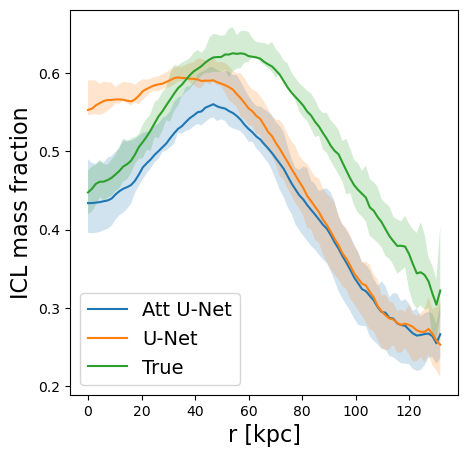}
    \caption{ICL mass fraction profiles of the true and recovered distributions as a function of radius in the maps. Shaded bands report the 16$^\mathrm{th}-$84$^\mathrm{th}$ percentiles.}
    \label{fig:all_profile}
\end{figure}

\subsection{Fiducial model}
Fig.~\ref{fig:result3} shows the results for two randomly chosen images from the test set. The colour bar guides the lecture: orange marks areas where the ICL fraction is 1 and black denotes 0. In each row, we present the expected output from {\tt ICL-SubFind} (left panel) which we take as the ground truth, the predictions by the Attention U-Net (central panel), and the original U-Net (right panel). We estimate the ICL mass fraction within each image.
\par
Most of the stellar mass belongs to the central galaxy and the substructures we mask. The visual inspection shows that both methods are generally able to recover the large-scale ICL structure of the halos. This is further confirmed by the ICL fraction reported in each panel, which does not significantly vary in these cases. Although the ICL distributions are not perfectly matched, cluster centres and masked interlopers (dark circles) are recovered. We observe that none of the DL models fully predicts the gradient in the ICL fraction at smaller radii, nevertheless, the Attention U-Net behaves significantly better in this metric: this is not surprising as the Attention U-Net is designed to reinforce learning in the regions of interest. However, we tested that a longer training phase only leads to overfitting and does not improve the result. Spurious elements (e.g. substructures not properly masked in the velocity dispersion) are also better classified by the Attention U-Net: an example is given on the lower right-hand side of the images in the lower row. Here, the Attention U-Net can minimise the contribution of ICL (effective null in the true distribution) whereas the U-Net predicts fractions close to 0.8. Clearly, masking galaxies will require using the same choice as in the analysis of observational data. We notice that neither the Attention U-Net nor the U-Net exhibits an ICL mass fraction equal to 1 (or 0) anywhere, which represents the case for many pixels in the true image. Although this result is unlikely to be physically meaningful, the networks lack this feature in the final mapping.
\par
\begin{figure}
    \centering
    \includegraphics[scale=0.6]{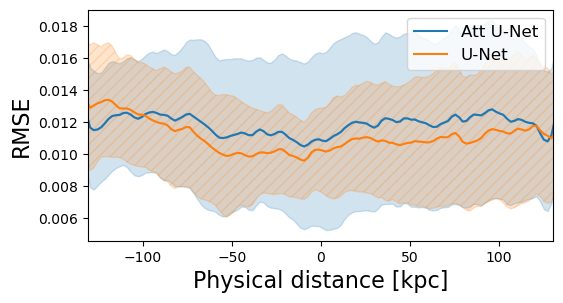}
    \caption{Standard deviation of the predictions from both networks (blue the Attention U-Net and orange the U-Net) integrated along the vertical axis. The distribution is smoothed within 4 kpc. The solid and dashed curves show the results for the two directions separately. The shaded areas mark the internal scatter in the sample. }
    \label{fig:MSE_pixel}
\end{figure}

For a more quantitative comparison, an evaluation of the median profiles of the ICL fraction in the true and recovered maps as a function of the distance -- as plotted in Fig.~\ref{fig:all_profile} -- can help us further acknowledge how significant these effects are on the general population of simulated clusters. We also report with shaded bands the 16$^\mathrm{th}-$84$^\mathrm{th}$ percentiles of the sample. The Attention U-Net behaves better than the U-Net, as it is mostly in agreement at the centre and tends on average to be lower than the true profile only at larger radii. 
This result is consistent with a systematic prediction for lower ICL fraction (by $\sim 20$\%) in the outskirts of the predicted maps, as we discussed. In any case, we find that the peak and the subsequent decrease of the ICL fraction are consistent with the ICL distribution, especially in the Attention U-Net case. The same conclusions can be drawn when inspecting the statistical errors from the predictions in the complete test sets. Rather than accounting for pixel-to-pixel misclassification (thereby excluding potential noise), we smooth the predicted and the true ICL distributions with a Gaussian filter over a physical scale of 4 kpc and we estimate a cumulative error as the standard deviation between the two smoothed images integrated along each direction (vertically and horizontally). This should allow us to appreciate where most of the differences in our predictions lie in the larger cluster environment. This result is presented in Fig.~\ref{fig:MSE_pixel} in terms of the root MSE (RMSE) which describes the average magnitude of errors in a regression model. Shaded regions report the internal scatter for both networks. The RMSE has an almost constant trend around $\sim 0.012$, although the Attention U-Net has a slightly larger scatter than the U-Net. Another way of visualising this error is by plotting the residual images between the true and predicted distribution as in Fig. \ref{fig:residuals_D1_26} for the halos in Fig. \ref{fig:result3}. The U-Net is more prone to errors at all scales suggesting that the model responds better when the learning is reinforced with Attention mechanisms.

\begin{figure*}
    \centering
    \includegraphics[scale=0.55]{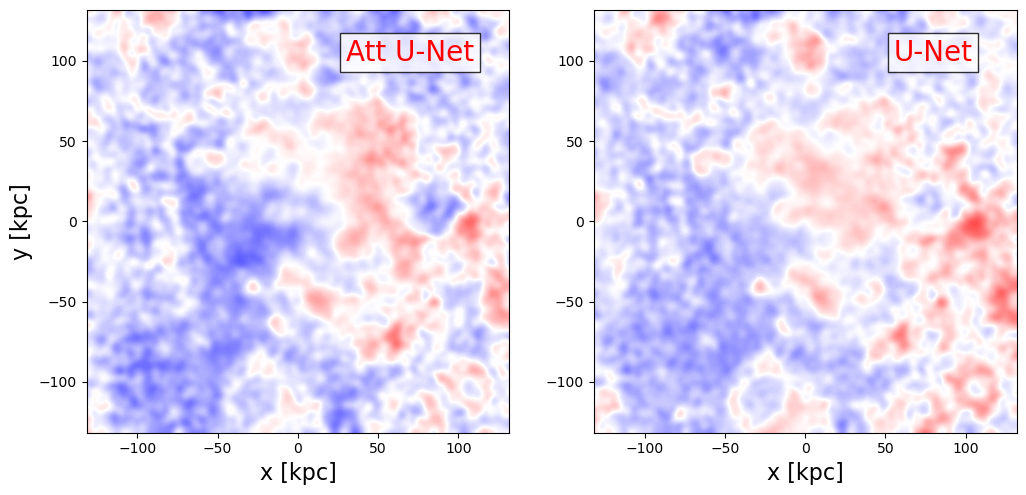}
\end{figure*}
\begin{figure*}
    \centering
    \includegraphics[scale=0.55]{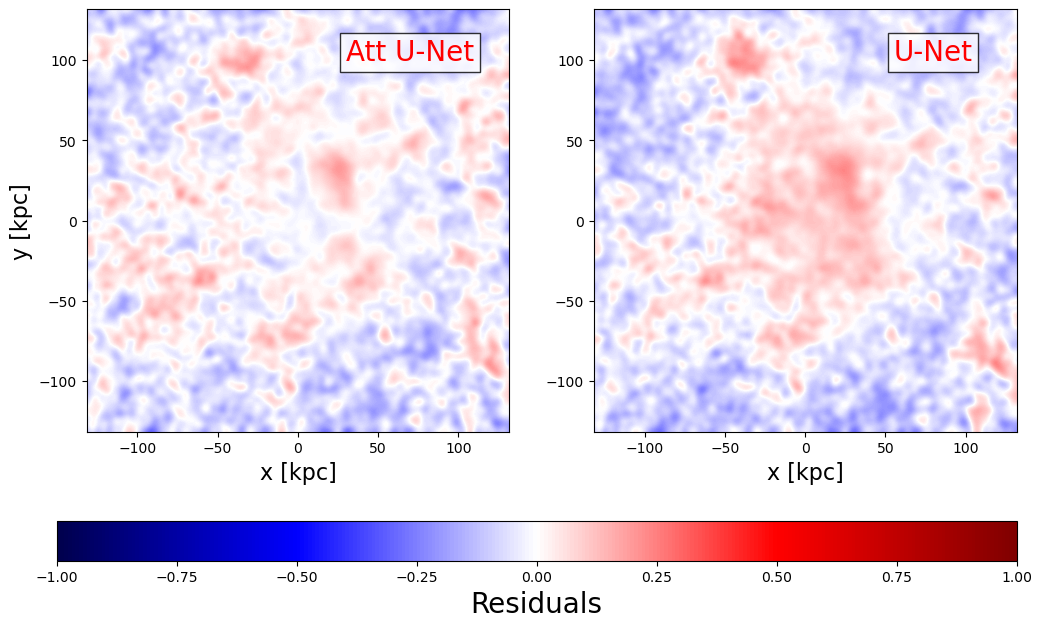}
    \caption{Residual heatmaps between the predicted and true ICL distribution for both models. The distribution is smoothed within 4 kpc. The corresponding ICL map distributions are plotted in Fig. \ref{fig:result3}.}
    \label{fig:residuals_D1_26}
\end{figure*}
\subsection{Limitations of the algorithms}
In the previous sections, we argued that the trained Attention U-Net predicts reasonable ICL distributions to the expected labels. Here, we discuss the foreseeable limitations of the algorithm and attempt to quantify such losses. 
\subsubsection{Size of the training set}
\label{subsec:size of the training set}
Generally, the training of DL models can suffer from the limited size of the training data because of overfitting or underfitting. Thus, it might be worth discussing this issue here, especially in our case where from a restricted number of halos we created our sample through data augmentation.
\par
\begin{figure}
    \centering
    \includegraphics[scale=0.72]{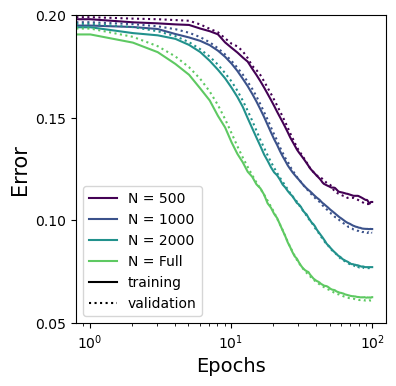}
    \caption{Learning curve of the model with increasing training set size under the same hyperparameters (i.e. 500, 1000, 2000, and all the examples). We show the training and validation error across different epochs down to the flattening of the learning curve. }
    \label{fig:sample_size}
\end{figure}

In Fig.~\ref{fig:sample_size} we propose the following experiment: we evaluate the learning curve of our model during the training phase with different sample sizes (i.e. 500, 1000, 2000 examples, and the full training set). In other words, for each epoch, we estimate the training and the validation errors as a function of the epoch. We remind that the training (validation) error is the error committed in the prediction over the training (validation) set as measured by the loss function (i.e. the MSE). It is remarkable to notice that in all cases we find a learning curve consistent with a well-trained network (i.e. a smooth decrease of both errors to reach a plateau). The plateau in the final phases of the training marks the minima of the gradient descent evolution of the trainable parameters of the network, thus in all cases we do not encounter overfitting or underfitting. Conversely, the errors involved in the process vary with the sample size, reaching their minimum for the full sample. From this, we conclude that the limitation on the data size can significantly impact the learning mechanism and probably represents one of the major limitations in our setup. On the other hand, progressing in the data augmentation operation to enlarge the training size can also negatively impact our training, as we might undergo overfitting considering the limited sample of objects. Having more independent clusters realisation would boost the learning. 

\subsubsection{IllustrisTNG}
We run a second test on the data, namely, we use the best model on a different set of simulations to qualitatively assess the impact of modelled physics in the predictions of ICL mass fractions. When assuming a definition of ICL based on dynamical behaviours many elements can affect the distribution of stellar particles and their kinematics, mostly related to the adopted sub-resolution models of star formation and energy feedback. A concrete example is provided by the AGN feedback which regulates star formation in massive galaxies, particularly impacting BCG masses \citep[e.g.][]{ragone-figueroa_brightest_2013}. 
\par
For this task, we select three of the most massive halos in IllustrisTNG-300 at our fiducial redshift (i.e. $z=0.3$). IllustrisTNG-300, corresponding to the largest cosmological box available in the IllustrisTNG suite \citep{pillepich_simulating_2018}, includes 75000$^3$ particles in a 302.6$^3 \, h^{-3}$ cMpc$^{3}$ volume. We chose this run as its mass and spatial resolutions are similar to the Dianoga set. The dark matter (initial gas) particles have mass $5.9\times 10^{7}\, M_{\odot}$ ($1.1\times 10^{7} \,M_{\odot}$) and Plummer equivalent softening $1.48$ kpc ($0.37$ ckpc). Details on the simulation and the astrophysical subgrid models implemented can be found in \citet{pillepich_first_2018}. A dynamical classification of ICL and BCG-bound stars has never been attempted on IllustrisTNG, thus we leave a complete analysis of this aspect to future work. Conversely, we blindly run the {\tt ICL-SubFind} algorithm to classify stars (see \ref{subsec:icl_identification}) in the selected halos and we visually inspect the properties of the two stellar components. We remark that in \cite{marini_machine_2022} we already discussed the performance of the random forest for simulations including different physical models and resolutions, being generally consistent with the expected results. 
\par
\begin{figure}
    \centering   
    \includegraphics[scale=0.6]{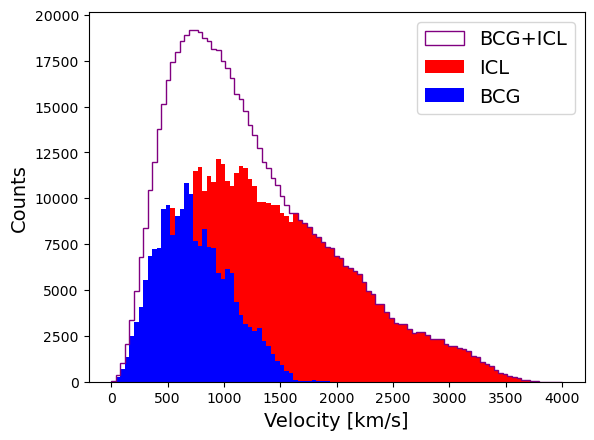}
    \caption{Distribution of the module of the three-dimensional velocity of the stars in the main halo selected among the IllustrisTNG sample. We plot the histograms for BCG (blue) and ICL (red) separately and combined (purple). }
    \label{fig:velocity_tng}
\end{figure}

In Fig.~\ref{fig:velocity_tng} we show the resulting distribution of star particle velocities in the two stellar components after running {\tt ICL-SubFind}. The purple histogram highlights the double-Maxwellian profile for BCG+ICL which can be traced back to the sum of the distributions of the two components. Thus, we can confirm that in IllustrisTNG we can also dynamically distinguish two stellar populations in the main halo associated with a central galaxy and a diffuse component.
\par
Based on this result, we carry out the same analysis (comprehensive of all the steps outlined in Sect. \ref{subsec:mock_images}). The results of this analysis are shown in Fig.~\ref{fig:tng}. The most important result is that the algorithm can spot the central galaxy position thanks to its kinematics and estimate its extension, as derived by the colour gradient in the ICL mass fraction distribution. Most interlopers are detected even though their size is often underestimated. Such error can lead to a systematic overestimation of the ICL stellar mass and points out the necessity of including different numerical simulations in future training. 
\par
Therefore, we conclude that optimal results are obtained for a set of images coming from our original set of simulations. Still, applying our Attention U-Net to mock images generated from a completely independent set of simulations also allows us to estimate the ICL and BCG extension happening at $\sim 50$ kpc from the centre. 

\begin{figure*}
    \centering
    \includegraphics[scale=0.5]{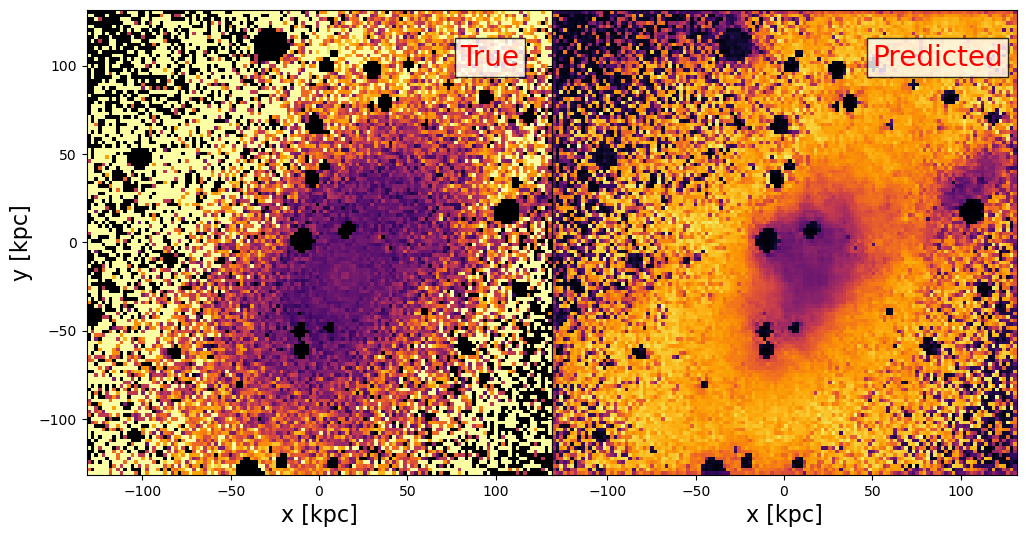}
\end{figure*}
\begin{figure*}
    \centering
    \includegraphics[scale=0.5]{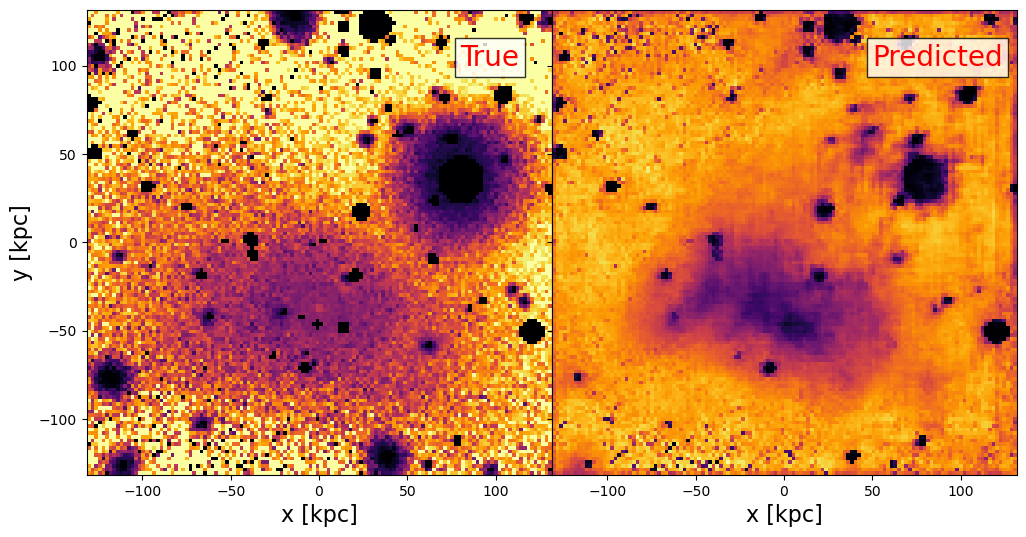}
\end{figure*}
\begin{figure*}
    \centering
    \includegraphics[scale=0.5]{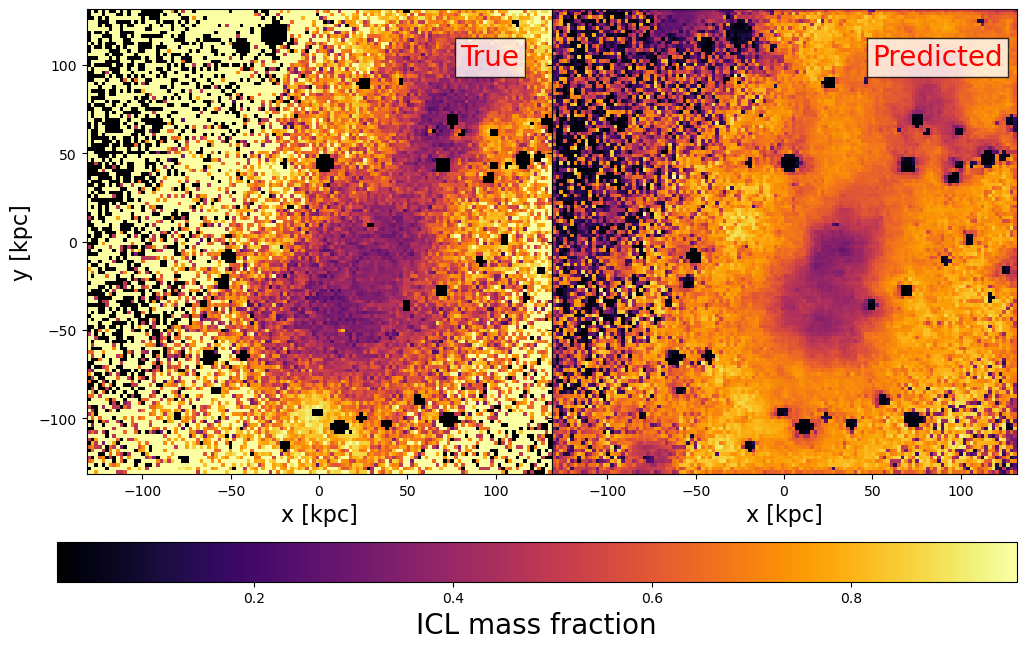}
    \caption{Distribution of the ICL mass fraction of the stars in the main halo of the three galaxy clusters selected within the IllustrisTNG simulations. The colour bar reports the higher ICL fractions in orange and black for lower. }
    \label{fig:tng}
\end{figure*}

\begin{figure*}
    \centering
    \subfigure[]{\includegraphics[trim={0 2.8cm 0 0},clip, width=0.33\textwidth]{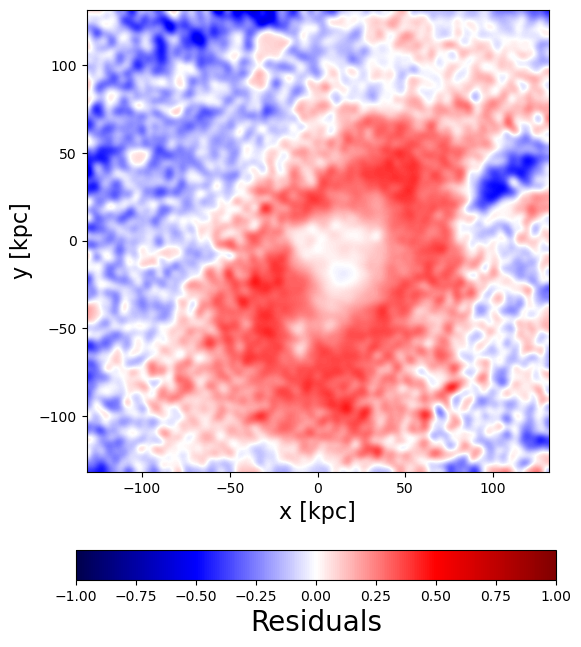}} 
    \subfigure[]{\includegraphics[trim={0 2.8cm 0 0},clip,width=0.33\textwidth]{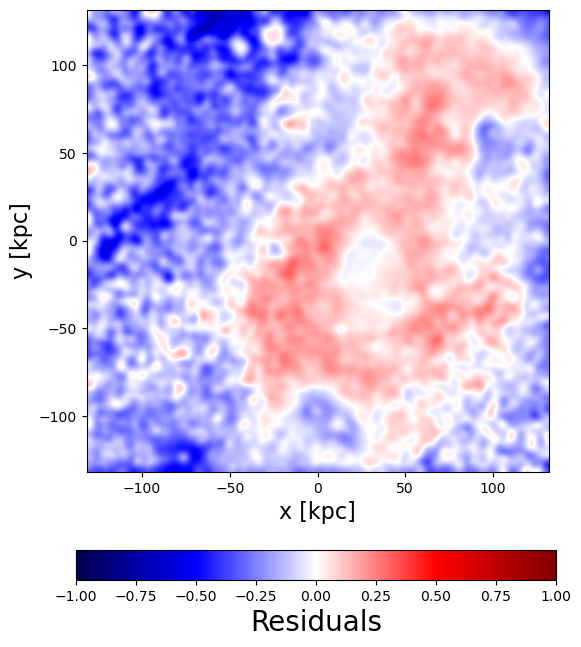}} 
    \subfigure[]{\includegraphics[trim={0 2.8cm 0 0},clip,width=0.33\textwidth]{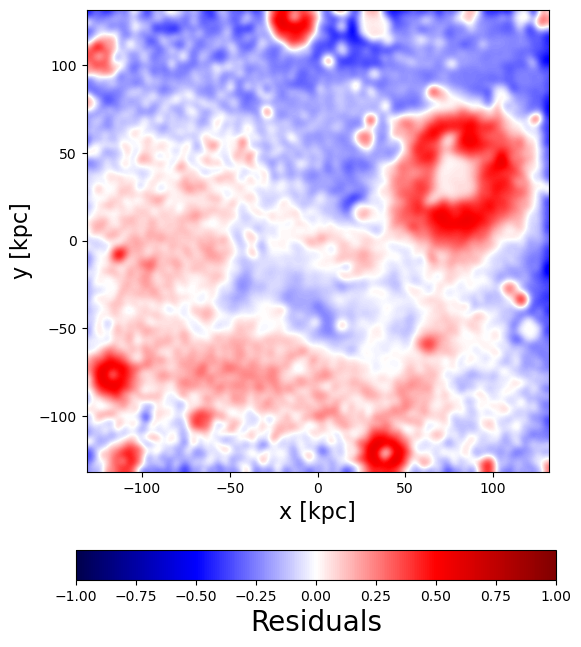}}
    \caption{Residual heatmaps between the predicted and true ICL distribution for the Attention U-Net applied to the IllustrisTNG cluster set reported in Fig.~\ref{fig:tng}. The distribution is smoothed within 4 kpc. The colour-bar follows Fig.~\ref{fig:residuals_D1_26}.}
    \label{fig:entropy}
\end{figure*}

 \section{Discussion and conclusions}
 \label{sec:conclusions}
 The challenge of identifying the ICL in galaxy clusters and groups is a long-standing issue. It has become a matter of increasing interest, as current and future surveys gather data of increasing statistics and sensitivity. One example is provided by the recent observations carried out by JWST, which is providing deep and high spatial resolution images to study ICL with a high signal-to-noise up to a radial distance of $\sim 400$ kpc \citep{montes_new_2022}, twice as far than previous HST studies \citep[e.g.][]{demaio_lost_2018}. This opens up the possibility of exploring the rich mixture of processes that drive the formation of the ICL.
 \par 
 In this paper, we discuss a new technique to separate the contributions of the BCG and the ICL to the distribution of stars in the central regions of galaxy clusters. We argue that the stellar velocity dispersion is an indirect tracer of the underlying population segregated in the galaxies and central halo. Thus, we can deduce the properties of this component through a kinematical decomposition. Observationally, one can determine the ICL distribution in a cluster after several assumptions on the light profile, defining the ICL to be the contribution to the faint light (i.e. below a given surface brightness limit) in clusters and groups of galaxies \citep[e.g.][]{mihos_burrell_2017,montes_intracluster_2019}; conversely, other studies \citep{kluge_structure_2020, spavone_fornax_2020} fit composite models to derive the stellar distribution in the light profile. The results of this latter approach are sensitive to the number of profiles fitted, each associated with a stellar component, and they might not be necessarily linked to the halo's assembly history \citep{remus_outer_2017}. Several authors have proposed a wavelet-like decomposition technique to extract the ICL from photometric images \citep{da_rocha_intragroup_2005,ellien_dawis_2021}. However, not only do such methods often provide different results according to the assumption chosen, but they also do not necessarily detect the dynamical differences expected to differentiate the ICL from the BCG stellar components. Here, we underline the importance of dynamically identifying ICL both in observational analyses and simulations. In principle, the scatter in the observationally inferred values of the ICL mass fraction, presented so far in the literature \citep[e.g.][]{rudick_quantity_2011, kluge_photometric_2021}, could be alleviated by standardising the methods used to define this component, although we are still far from having settled the matter.
 \par Cosmological hydrodynamical simulations suggest that the analysis of stellar kinematics offers a physically grounded approach to define the separation between BCG and ICL components \citep[e.g.][]{dolag_dynamical_2010, marini_machine_2022}. However, detailed spectroscopic observations to characterise the kinematics of the ICL are limited by the low surface brightness regime. Most spectroscopic data are restricted to 3$-$4 BCG effective radii (e.g. recently \citealt{boardman_integral-field_2017, loubser_merger_2022}). At larger radii, discrete tracers are extensively used \citep[(e.g. planetary nebulae and globular clusters; see][for a review]{arnaboldi_kinematics_2022}. Ideally, recovering the kinematics of the stellar component would help us decipher the variety of results coming from different methods and homogenise the results in the literature.
 \par Our work has shown that we can train a network to infer the ICL distributions from stellar kinematics. Thus, we can adapt and refine this method (for example, using different cosmological simulations to reduce the dependence on the numerical schemes) to suit our needs in future observations, providing a physically motivated distinction between the ICL and the central galaxy. 
\par In this study, we selected a sample of clusters (among the most massive) from the Dianoga zoom-in cosmological hydrodynamical simulations, which we used to create mock images of galaxy clusters. From the original dataset, we performed data augmentation by modifying (e.g. resizing, zooming, rotating) the images and included a simplified treatment of contaminants (i.e. interlopers) by masking substructures in the image. The final dataset is composed of line-of-sight velocity dispersion and ICL mass fraction maps divided into training, validation and test sets. Finally, we trained two models (U-Net and Attention U-Net, \citealt{ronneberger_u-net_2015}) with the One-Cycle policy \citep{smith_super-convergence_2018} and we discussed the accuracy of the networks. Our results can be summarised as follows. \begin{itemize} 
\item By modelling the dynamical separation of the BCG and ICL stellar components in our simulations \citep[e.g.][]{dolag_dynamical_2010,marini_machine_2022}, cluster-size and group-size halos have a transition radius at roughly $0.1\, R_{200}$, corresponding to the region where the ICL dominates the stellar component. We expect that covering with observations the scales corresponding to this transition radius should be effective in tracing the dynamical features of the ICL in the stellar kinematics (see Fig.~\ref{fig:10percR200}). 
\item The application of both U-Net and Attention U-Net models to mock velocity dispersion maps allows us to recover the large-scale ICL structure, though some gradients and small-scale features are not fully captured. The Attention U-Net generally outperforms the U-Net, particularly in capturing smaller-scale features and reducing spurious elements, as highlighted in Fig.~\ref{fig:all_profile}. 
\item The most significant limitations to our model are connected to the limited data size -- as described in Fig.~\ref{fig:sample_size} -- and the unique numerical scheme included in the training of the network. Such conditions impact the accuracy of the model predictions, as proved when run on halos from the IllustrisTNG-300 -- see Fig.~\ref{fig:tng}. A viable solution would be to expand the training phase to include a set of simulations more extended in terms of the number of clusters simulated, implementations of the relevant physical processes driving galaxy formation, and numerical resolution. 
\end{itemize}
In conclusion, the method presented here proved sufficiently reliable in characterising the ICL distribution in our simulations set only from the projected phase-space information. As a final remark, we shall refrain from claiming that the network will perform at this level of accuracy for real spectroscopic observations or other simulations, as its accuracy has shown to be dependent on the details of the numerical implementation of the physical processes included in the training data. In this sense, a natural follow-up would be the design of a training set effectively mocking the observational conditions of an IFS (e.g. spectral features and signal-to-noise ratio) to guide observers in their task. Furthermore, the analysis should be extended to investigate the role of the halo's dynamical state (i.e. relaxed or disturbed) in affecting the recovery rate of the maps, since disruptive events can impact both the stellar kinematics \citep{longobardi_build-up_2015} and ICL fraction \citep{contini_intracluster_2023}. On the other hand, this method paves the way for CNNs as powerful tools for constructing a robust pipeline of ICL detection, taking advantage of high-sensitivity spectroscopic studies of stellar kinematics in central regions of clusters and groups.

\begin{acknowledgements}
      We thank Amelia Fraser-McKelvie for her support on the technical aspects of the observation. We are grateful to Magda Arnaboldi, Claudia Pulsoni, and Paola Popesso for the useful discussions that led to significant improvements to the paper's draft. We acknowledge the CINECA award under the ISCRA initiative, for the availability of high-performance computing resources and support. IM acknowledges support from the European Research Council (ERC) under the European Union’s Horizon Europe research and innovation programme ERC CoG (Grant agreement No. 101045437, PI P. Popesso). Simulations have been carried out: using MARCONI at CINECA (Italy), with CPU time assigned through grants ISCRA B, and through INAF-CINECA and University of Trieste – CINECA agreements; at the Tianhe-2 platform of the Guangzhou Supercomputer Center by the support from the National Key Program for Science and Technology Research and Development (2017YFB0203300). This paper is supported by the Fondazione ICSC National Recovery and Resilience Plan (PNRR) Project ID CN-00000013 "Italian Research Center on High-Performance Computing, Big Data and Quantum Computing" funded by MUR Missione 4 Componente 2 Investimento 1.4: "Potenziamento strutture di ricerca e creazione di "campioni nazionali di R$\&$S (M4C2-19 )" - Next Generation EU (NGEU). SB acknowledges partial financial support from the INFN Indark Grant.
\end{acknowledgements}

   \bibliographystyle{aa} % style aa.bst
   \bibliography{references} % your references Yourfile.bib

\begin{thebibliography}{91}
\expandafter\ifx\csname natexlab\endcsname\relax\def\natexlab#1{#1}\fi

\bibitem[{Agarap(2019)}]{agarap_deep_2019}
Agarap, A.~F. 2019

\bibitem[{Alonso Asensio {et~al.}(2020)Alonso Asensio, Dalla Vecchia, Bahé,
  Barnes, \& Kay}]{alonsoasensio_intracluster_2020}
Alonso Asensio, I., Dalla Vecchia, C., Bahé, Y.~M., Barnes, D.~J., \& Kay,
  S.~T. 2020, MNRAS, 494, 1859

\bibitem[{Arnaboldi \& Gerhard(2022)}]{arnaboldi_kinematics_2022}
Arnaboldi, M. \& Gerhard, O. 2022, Frontiers in Astronomy and Space Sciences, 9

\bibitem[{Arnaboldi \& Napolitano(2001)}]{arnaboldi_dynamics_2001}
Arnaboldi, M. \& Napolitano, N.~R. 2001, 230, 409

\bibitem[{Bacon {et~al.}(2010)Bacon, Accardo, Adjali, Anwand, Bauer, Biswas,
  Blaizot, Boudon, Brau-Nogue, Brinchmann, Caillier, Capoani, Carollo, Contini,
  Couderc, Daguisé, Deiries, Delabre, Dreizler, Dubois, Dupieux, Dupuy,
  Emsellem, Fechner, Fleischmann, François, Gallou, Gharsa, Glindemann, Gojak,
  Guiderdoni, Hansali, Hahn, Jarno, Kelz, Koehler, Kosmalski, Laurent,
  Le~Floch, Lilly, Lizon, Loupias, Manescau, Monstein, Nicklas, Olaya, Pares,
  Pasquini, Pécontal-Rousset, Pelló, Petit, Popow, Reiss, Remillieux,
  Renault, Roth, Rupprecht, Serre, Schaye, Soucail, Steinmetz, Streicher,
  Stuik, Valentin, Vernet, Weilbacher, Wisotzki, \& Yerle}]{bacon_muse_2010}
Bacon, R., Accardo, M., Adjali, L., {et~al.} 2010, 7735, 773508

\bibitem[{Ball {et~al.}(2004)Ball, Loveday, Fukugita, Nakamura, Okamura,
  Brinkmann, \& Brunner}]{ball_galaxy_2004}
Ball, N.~M., Loveday, J., Fukugita, M., {et~al.} 2004, MNRAS, 348, 1038

\bibitem[{Banerji {et~al.}(2010)Banerji, Lahav, Lintott, Abdalla, Schawinski,
  Bamford, Andreescu, Murray, Raddick, Slosar, Szalay, Thomas, \&
  Vandenberg}]{banerji_galaxy_2010}
Banerji, M., Lahav, O., Lintott, C.~J., {et~al.} 2010, MNRAS, 406, 342

\bibitem[{Bassini {et~al.}(2020)Bassini, Rasia, Borgani, Granato,
  Ragone-Figueroa, Biffi, Ragagnin, Dolag, Lin, Murante, Napolitano, Taffoni,
  Tornatore, \& Wang}]{bassini_dianoga_2020}
Bassini, L., Rasia, E., Borgani, S., {et~al.} 2020, A\&A, 642, A37

\bibitem[{Beck {et~al.}(2016)Beck, Murante, Arth, Remus, Teklu, Donnert,
  Planelles, Beck, Förster, Imgrund, Dolag, \& Borgani}]{beck_improved_2016}
Beck, A.~M., Murante, G., Arth, A., {et~al.} 2016, MNRAS, 455, 2110

\bibitem[{Binney \& Tremaine(2011)}]{binney_galactic_2011}
Binney, J. \& Tremaine, S. 2011, Galactic {Dynamics}: {Second} {Edition}
  (Princeton University Press)

\bibitem[{Boardman {et~al.}(2017)Boardman, Weijmans, van~den Bosch, Kuntschner,
  Emsellem, Cappellari, de~Zeeuw, Falcón-Barroso, Krajnović, McDermid, Naab,
  van~de Ven, \& Yildirim}]{boardman_integral-field_2017}
Boardman, N.~F., Weijmans, A.-M., van~den Bosch, R., {et~al.} 2017, MNRAS, 471,
  4005

\bibitem[{Burke {et~al.}(2019)Burke, Aleo, Chen, Liu, Peterson, Sembroski, \&
  Lin}]{burke_deblending_2019}
Burke, C.~J., Aleo, P.~D., Chen, Y.-C., {et~al.} 2019, MNRAS, 490, 3952

\bibitem[{Bílek {et~al.}(2020)Bílek, Duc, Cuillandre, Gwyn, Cappellari,
  Bekaert, Bonfini, Bitsakis, Paudel, Krajnović, Durrell, \&
  Marleau}]{bilek_census_2020}
Bílek, M., Duc, P.-A., Cuillandre, J.-C., {et~al.} 2020, MNRAS, 498, 2138

\bibitem[{Cantat-Gaudin {et~al.}(2020)Cantat-Gaudin, Anders, Castro-Ginard,
  Jordi, Romero-Gómez, Soubiran, Casamiquela, Tarricq, Moitinho, Vallenari,
  Bragaglia, Krone-Martins, \& Kounkel}]{cantat-gaudin_painting_2020}
Cantat-Gaudin, T., Anders, F., Castro-Ginard, A., {et~al.} 2020, A\&A, 640, A1

\bibitem[{Carleo {et~al.}(2019)Carleo, Cirac, Cranmer, Daudet, Schuld, Tishby,
  Vogt-Maranto, \& Zdeborová}]{carleo_machine_2019}
Carleo, G., Cirac, I., Cranmer, K., {et~al.} 2019, Reviews of Modern Physics,
  91, 045002

\bibitem[{Chadayammuri {et~al.}(2023)Chadayammuri, Ntampaka, ZuHone,
  Bogd{\'a}n, \& Kraft}]{chadayammuri_painting_2023}
Chadayammuri, U., Ntampaka, M., ZuHone, J., Bogd{\'a}n, {\'A}., \& Kraft, R.~P.
  2023, MNRAS, 526, 2812

\bibitem[{Chen {et~al.}(2022)Chen, Zu, Shao, \& Shan}]{chen_sphere_2022}
Chen, X., Zu, Y., Shao, Z., \& Shan, H. 2022, MNRAS, 514, 2692

\bibitem[{Collister \& Lahav(2004)}]{collister_annz_2004}
Collister, A.~A. \& Lahav, O. 2004, Publications of the Astronomical Society of
  the Pacific, 116, 345

\bibitem[{Contini(2021)}]{contini_origin_2021}
Contini, E. 2021, Galaxies, 9, 60

\bibitem[{Contini {et~al.}(2022)Contini, Chen, \& Gu}]{contini_transition_2022}
Contini, E., Chen, H.~Z., \& Gu, Q. 2022, ApJ, 928, 99

\bibitem[{Contini \& Gu(2021)}]{contini_brightest_2021}
Contini, E. \& Gu, Q. 2021, ApJ, 915, 106

\bibitem[{Contini {et~al.}(2023)Contini, Jeon, Rhee, Han, \&
  Yi}]{contini_intracluster_2023}
Contini, E., Jeon, S., Rhee, J., Han, S., \& Yi, S.~K. 2023, ApJ, 958, 72

\bibitem[{Da~Rocha \& Mendes~de Oliveira(2005)}]{da_rocha_intragroup_2005}
Da~Rocha, C. \& Mendes~de Oliveira, C. 2005, MNRAS, 364, 1069

\bibitem[{DeMaio {et~al.}(2018)DeMaio, Gonzalez, Zabludoff, Zaritsky, Connor,
  Donahue, \& Mulchaey}]{demaio_lost_2018}
DeMaio, T., Gonzalez, A.~H., Zabludoff, A., {et~al.} 2018, MNRAS, 474, 3009

\bibitem[{Dieleman {et~al.}(2015)Dieleman, Willett, \&
  Dambre}]{dieleman_rotation-invariant_2015}
Dieleman, S., Willett, K.~W., \& Dambre, J. 2015, MNRAS, 450, 1441

\bibitem[{Dolag {et~al.}(2009)Dolag, Borgani, Murante, \&
  Springel}]{dolag_substructures_2009}
Dolag, K., Borgani, S., Murante, G., \& Springel, V. 2009, MNRAS, 399, 497

\bibitem[{Dolag {et~al.}(2010)Dolag, Murante, \&
  Borgani}]{dolag_dynamical_2010}
Dolag, K., Murante, G., \& Borgani, S. 2010, MNRAS, 405, 1544

\bibitem[{Dolag {et~al.}(2005)Dolag, Vazza, Brunetti, \&
  Tormen}]{dolag_turbulent_2005}
Dolag, K., Vazza, F., Brunetti, G., \& Tormen, G. 2005, MNRAS, 364, 753

\bibitem[{Domínguez~Sánchez {et~al.}(2018)Domínguez~Sánchez,
  Huertas-Company, Bernardi, Tuccillo, \&
  Fischer}]{dominguez_sanchez_improving_2018}
Domínguez~Sánchez, H., Huertas-Company, M., Bernardi, M., Tuccillo, D., \&
  Fischer, J.~L. 2018, MNRAS, 476, 3661

\bibitem[{Ellien {et~al.}(2021)Ellien, Slezak, Martinet, Durret, Adami,
  Gavazzi, Rabaça, Da~Rocha, \& Epitácio~Pereira}]{ellien_dawis_2021}
Ellien, A., Slezak, E., Martinet, N., {et~al.} 2021, A\&A, 649, A38

\bibitem[{Feldmann {et~al.}(2006)Feldmann, Carollo, Porciani, Lilly, Capak,
  Taniguchi, Le~Fèvre, Renzini, Scoville, Ajiki, Aussel, Contini, McCracken,
  Mobasher, Murayama, Sanders, Sasaki, Scarlata, Scodeggio, Shioya, Silverman,
  Takahashi, Thompson, \& Zamorani}]{feldmann_zurich_2006}
Feldmann, R., Carollo, C.~M., Porciani, C., {et~al.} 2006, MNRAS, 372, 565

\bibitem[{George \& Huerta(2018)}]{george_deep_2018}
George, D. \& Huerta, E.~A. 2018, Physics Letters B, 778, 64

\bibitem[{Gibson {et~al.}(2012)Gibson, Aigrain, Roberts, Evans, Osborne, \&
  Pont}]{gibson_gaussian_2012}
Gibson, N.~P., Aigrain, S., Roberts, S., {et~al.} 2012, MNRAS, 419, 2683

\bibitem[{Glorot {et~al.}(2011)Glorot, Bordes, \& Bengio}]{glorot_deep_2011}
Glorot, X., Bordes, A., \& Bengio, Y. 2011, in Proceedings of the {Fourteenth}
  {International} {Conference} on {Artificial} {Intelligence} and {Statistics}
  (JMLR Workshop and Conference Proceedings), 315--323, iSSN: 1938-7228

\bibitem[{Gonzalez {et~al.}(2021)Gonzalez, George, Connor, Deason, Donahue,
  Montes, Zabludoff, \& Zaritsky}]{gonzalez_discovery_2021}
Gonzalez, A.~H., George, T., Connor, T., {et~al.} 2021, MNRAS, 507, 963

\bibitem[{Gonzalez {et~al.}(2007)Gonzalez, Zaritsky, \&
  Zabludoff}]{gonzalez_census_2007}
Gonzalez, A.~H., Zaritsky, D., \& Zabludoff, A.~I. 2007, ApJ, 666, 147

\bibitem[{Goodfellow {et~al.}(2016)Goodfellow, Bengio, \&
  Courville}]{goodfellow_deep_2016}
Goodfellow, I., Bengio, Y., \& Courville, A. 2016, Deep {Learning} (MIT Press)

\bibitem[{He {et~al.}(2015)He, Zhang, Ren, \& Sun}]{he_deep_2015}
He, K., Zhang, X., Ren, S., \& Sun, J. 2015, Deep {Residual} {Learning} for
  {Image} {Recognition}

\bibitem[{Jaffe(1983)}]{jaffe_simple_1983}
Jaffe, W. 1983, MNRAS, 202, 995

\bibitem[{Jetley {et~al.}(2018)Jetley, Lord, Lee, \& Torr}]{jetley_learn_2018}
Jetley, S., Lord, N.~A., Lee, N., \& Torr, P. H.~S. 2018, Learn {To} {Pay}
  {Attention}

\bibitem[{Kamdar {et~al.}(2016)Kamdar, Turk, \& Brunner}]{kamdar_machine_2016}
Kamdar, H.~M., Turk, M.~J., \& Brunner, R.~J. 2016, MNRAS, 457, 1162

\bibitem[{Karademir {et~al.}(2019)Karademir, Remus, Burkert, Dolag, Hoffmann,
  Moster, Steinwandel, \& Zhang}]{karademir_outer_2019}
Karademir, G.~S., Remus, R.-S., Burkert, A., {et~al.} 2019, MNRAS, 487, 318

\bibitem[{Kingma \& Ba(2017)}]{kingma_adam_2017}
Kingma, D.~P. \& Ba, J. 2017, Adam: {A} {Method} for {Stochastic}
  {Optimization}

\bibitem[{Kluge {et~al.}(2021)Kluge, Bender, Riffeser, Goessl, Hopp, Schmidt,
  \& Ries}]{kluge_photometric_2021}
Kluge, M., Bender, R., Riffeser, A., {et~al.} 2021, ApJS, 252, 27

\bibitem[{Kluge {et~al.}(2020)Kluge, Neureiter, Riffeser, Bender, Goessl, Hopp,
  Schmidt, Ries, \& Brosch}]{kluge_structure_2020}
Kluge, M., Neureiter, B., Riffeser, A., {et~al.} 2020, ApJS, 247, 43

\bibitem[{Kravtsov \& Borgani(2012)}]{kravtsov_formation_2012}
Kravtsov, A.~V. \& Borgani, S. 2012, Annu. Rev. Astron. Astrophys., 50, 353

\bibitem[{Krizhevsky {et~al.}(2012)Krizhevsky, Sutskever, \&
  Hinton}]{krizhevsky_imagenet_2012}
Krizhevsky, A., Sutskever, I., \& Hinton, G.~E. 2012, in Advances in {Neural}
  {Information} {Processing} {Systems}, Vol.~25 (Curran Associates, Inc.)

\bibitem[{LeCun {et~al.}(2015)LeCun, Bengio, \& Hinton}]{lecun_deep_2015}
LeCun, Y., Bengio, Y., \& Hinton, G. 2015, Nature, 521, 436

\bibitem[{LeCun {et~al.}(1998)LeCun, Bottou, Bengio, \&
  Haffner}]{lecun_gradient-based_1998}
LeCun, Y., Bottou, L., Bengio, Y., \& Haffner, P. 1998, Proceedings of the
  IEEE, 86, 2278

\bibitem[{Longobardi {et~al.}(2015)Longobardi, Arnaboldi, Gerhard, \&
  Mihos}]{longobardi_build-up_2015}
Longobardi, A., Arnaboldi, M., Gerhard, O., \& Mihos, J.~C. 2015, A\&A, 579, L3

\bibitem[{Loubser {et~al.}(2022)Loubser, Lagos, Babul, O’Sullivan, Jung,
  Olivares, \& Kolokythas}]{loubser_merger_2022}
Loubser, S.~I., Lagos, P., Babul, A., {et~al.} 2022, MNRAS, 515, 1104

\bibitem[{Marini {et~al.}(2021)Marini, Borgani, Saro, Granato, Ragone-Figueroa,
  Sartoris, Dolag, Murante, Ragagnin, \& Wang}]{marini_velocity_2021}
Marini, I., Borgani, S., Saro, A., {et~al.} 2021, MNRAS, 507, 5780

\bibitem[{Marini {et~al.}(2022)Marini, Borgani, Saro, Murante, Granato,
  Ragone-Figueroa, \& Taffoni}]{marini_machine_2022}
Marini, I., Borgani, S., Saro, A., {et~al.} 2022, MNRAS, 514, 3082

\bibitem[{Mihos {et~al.}(2017)Mihos, Harding, Feldmeier, Rudick, Janowiecki,
  Morrison, Slater, \& Watkins}]{mihos_burrell_2017}
Mihos, J.~C., Harding, P., Feldmeier, J.~J., {et~al.} 2017, ApJ, 834, 16

\bibitem[{Montenegro-Taborda {et~al.}(2023)Montenegro-Taborda, Rodriguez-Gomez,
  Pillepich, Avila-Reese, Sales, Rodríguez-Puebla, \&
  Hernquist}]{montenegro-taborda_growth_2023}
Montenegro-Taborda, D., Rodriguez-Gomez, V., Pillepich, A., {et~al.} 2023,
  MNRAS, 521, 800

\bibitem[{Montes(2022)}]{montes_faint_2022}
Montes, M. 2022, Nature Astronomy, 6, 308

\bibitem[{Montes {et~al.}(2021)Montes, Brough, Owers, \&
  Santucci}]{montes_buildup_2021}
Montes, M., Brough, S., Owers, M.~S., \& Santucci, G. 2021, ApJ, 910, 45

\bibitem[{Montes \& Trujillo(2019)}]{montes_intracluster_2019}
Montes, M. \& Trujillo, I. 2019, MNRAS, 482, 2838

\bibitem[{Montes \& Trujillo(2022)}]{montes_new_2022}
Montes, M. \& Trujillo, I. 2022, ApJ, 940, L51

\bibitem[{Murante {et~al.}(2004)Murante, Arnaboldi, Gerhard, Borgani, Cheng,
  Diaferio, Dolag, Moscardini, Tormen, Tornatore, \&
  Tozzi}]{murante_diffuse_2004}
Murante, G., Arnaboldi, M., Gerhard, O., {et~al.} 2004, ApJ, 607, L83

\bibitem[{Navarro {et~al.}(1997)Navarro, Frenk, \&
  White}]{navarro_universal_1997}
Navarro, J.~F., Frenk, C.~S., \& White, S. D.~M. 1997, ApJ, 490, 493

\bibitem[{Oktay {et~al.}(2018)Oktay, Schlemper, Folgoc, Lee, Heinrich, Misawa,
  Mori, McDonagh, Hammerla, Kainz, Glocker, \& Rueckert}]{oktay_attention_2018}
Oktay, O., Schlemper, J., Folgoc, L.~L., {et~al.} 2018

\bibitem[{Paszke {et~al.}(2019)Paszke, Gross, Massa, Lerer, Bradbury, Chanan,
  Killeen, Lin, Gimelshein, Antiga, Desmaison, Köpf, Yang, DeVito, Raison,
  Tejani, Chilamkurthy, Steiner, Fang, Bai, \& Chintala}]{paszke_pytorch_2019}
Paszke, A., Gross, S., Massa, F., {et~al.} 2019

\bibitem[{Pillepich {et~al.}(2018{\natexlab{a}})Pillepich, Nelson, Hernquist,
  Springel, Pakmor, Torrey, Weinberger, Genel, Naiman, Marinacci, \&
  Vogelsberger}]{pillepich_first_2018}
Pillepich, A., Nelson, D., Hernquist, L., {et~al.} 2018{\natexlab{a}}, MNRAS,
  475, 648

\bibitem[{Pillepich {et~al.}(2018{\natexlab{b}})Pillepich, Springel, Nelson,
  Genel, Naiman, Pakmor, Hernquist, Torrey, Vogelsberger, Weinberger, \&
  Marinacci}]{pillepich_simulating_2018}
Pillepich, A., Springel, V., Nelson, D., {et~al.} 2018{\natexlab{b}}, MNRAS,
  473, 4077

\bibitem[{Pop {et~al.}(2018)Pop, Pillepich, Amorisco, \&
  Hernquist}]{pop_formation_2018}
Pop, A.-R., Pillepich, A., Amorisco, N.~C., \& Hernquist, L. 2018, MNRAS, 480,
  1715

\bibitem[{Proctor {et~al.}(2023)Proctor, Lagos, Ludlow, \&
  Robotham}]{proctor_identifying_2023}
Proctor, K.~L., Lagos, C. d.~P., Ludlow, A.~D., \& Robotham, A. S.~G. 2023

\bibitem[{Ragone-Figueroa {et~al.}(2018)Ragone-Figueroa, Granato, Ferraro,
  Murante, Biffi, Borgani, Planelles, \& Rasia}]{ragone-figueroa_bcg_2018}
Ragone-Figueroa, C., Granato, G.~L., Ferraro, M.~E., {et~al.} 2018, MNRAS, 479,
  1125

\bibitem[{Ragone-Figueroa {et~al.}(2013)Ragone-Figueroa, Granato, Murante,
  Borgani, \& Cui}]{ragone-figueroa_brightest_2013}
Ragone-Figueroa, C., Granato, G.~L., Murante, G., Borgani, S., \& Cui, W. 2013,
  MNRAS, 436, 1750

\bibitem[{Remus {et~al.}(2017)Remus, Dolag, \& Hoffmann}]{remus_outer_2017}
Remus, R.-S., Dolag, K., \& Hoffmann, T. 2017, Galaxies, 5, 49

\bibitem[{Ronneberger {et~al.}(2015)Ronneberger, Fischer, \&
  Brox}]{ronneberger_u-net_2015}
Ronneberger, O., Fischer, P., \& Brox, T. 2015, U-{Net}: {Convolutional}
  {Networks} for {Biomedical} {Image} {Segmentation}

\bibitem[{Rudick {et~al.}(2011)Rudick, Mihos, \&
  McBride}]{rudick_quantity_2011}
Rudick, C.~S., Mihos, J.~C., \& McBride, C.~K. 2011, ApJ, 732, 48

\bibitem[{Salvato {et~al.}(2011)Salvato, Ilbert, Hasinger, Rau, Civano,
  Zamorani, Brusa, Elvis, Vignali, Aussel, Comastri, Fiore, Le~Floc'h,
  Mainieri, Bardelli, Bolzonella, Bongiorno, Capak, Caputi, Cappelluti,
  Carollo, Contini, Garilli, Iovino, Fotopoulou, Fruscione, Gilli, Halliday,
  Kneib, Kakazu, Kartaltepe, Koekemoer, Kovac, Ideue, Ikeda, Impey, Le~Fevre,
  Lamareille, Lanzuisi, Le~Borgne, Le~Brun, Lilly, Maier, Manohar, Masters,
  McCracken, Messias, Mignoli, Mobasher, Nagao, Pello, Puccetti, Perez-Montero,
  Renzini, Sargent, Sanders, Scodeggio, Scoville, Shopbell, Silvermann,
  Taniguchi, Tasca, Tresse, Trump, \& Zucca}]{salvato_dissecting_2011}
Salvato, M., Ilbert, O., Hasinger, G., {et~al.} 2011, ApJ, 742, 61

\bibitem[{Schanche {et~al.}(2019)Schanche, Collier~Cameron, Hébrard, Nielsen,
  Triaud, Almenara, Alsubai, Anderson, Armstrong, Barros, Bouchy, Boumis,
  Brown, Faedi, Hay, Hebb, Kiefer, Mancini, Maxted, Palle, Pollacco, Queloz,
  Smalley, Udry, West, \& Wheatley}]{schanche_machine-learning_2019}
Schanche, N., Collier~Cameron, A., Hébrard, G., {et~al.} 2019, MNRAS, 483,
  5534

\bibitem[{Schlemper {et~al.}(2019)Schlemper, Oktay, Schaap, Heinrich, Kainz,
  Glocker, \& Rueckert}]{schlemper_attention_2019}
Schlemper, J., Oktay, O., Schaap, M., {et~al.} 2019, Medical Image Analysis,
  53, 197

\bibitem[{Siddique {et~al.}(2021)Siddique, Paheding, Alom, \&
  Devabhaktuni}]{siddique_recurrent_2021}
Siddique, N., Paheding, S., Alom, M.~Z., \& Devabhaktuni, V. 2021, in Pattern
  {Recognition} and {Tracking} {XXXII}. {Vol}. 11735. {SPIE}, Vol. 11735,
  117350L

\bibitem[{Smith(2017)}]{smith_cyclical_2017}
Smith, L.~N. 2017, Cyclical {Learning} {Rates} for {Training} {Neural}
  {Networks}

\bibitem[{Smith \& Topin(2018)}]{smith_super-convergence_2018}
Smith, L.~N. \& Topin, N. 2018, Super-{Convergence}: {Very} {Fast} {Training}
  of {Neural} {Networks} {Using} {Large} {Learning} {Rates}

\bibitem[{Smith \& Geach(2023)}]{smith_astronomia_2023}
Smith, M.~J. \& Geach, J.~E. 2023, Royal Society Open Science, 10, 221454

\bibitem[{Spavone {et~al.}(2020)Spavone, Iodice, van~de Ven, Falcón-Barroso,
  Raj, Hilker, Peletier, Capaccioli, Mieske, Venhola, Napolitano, Cantiello,
  Paolillo, \& Schipani}]{spavone_fornax_2020}
Spavone, M., Iodice, E., van~de Ven, G., {et~al.} 2020, A\&A, 639, A14

\bibitem[{Springel(2005)}]{springel_cosmological_2005}
Springel, V. 2005, MNRAS, 364, 1105

\bibitem[{Springel \& Hernquist(2003)}]{springel_cosmological_2003}
Springel, V. \& Hernquist, L. 2003, MNRAS, 339, 289

\bibitem[{Springel {et~al.}(2001)Springel, White, Tormen, \&
  Kauffmann}]{springel_populating_2001}
Springel, V., White, S. D.~M., Tormen, G., \& Kauffmann, G. 2001, MNRAS, 328,
  726

\bibitem[{Sérsic(1963)}]{sersic_influence_1963}
Sérsic, J.~L. 1963, Boletin de la Asociacion Argentina de Astronomia La Plata
  Argentina, 6, 41

\bibitem[{Tormen \& Bertschinger(1996)}]{tormen_adding_1996}
Tormen, G. \& Bertschinger, E. 1996, ApJ, 472, 14

\bibitem[{Tornatore {et~al.}(2007)Tornatore, Borgani, Dolag, \&
  Matteucci}]{tornatore_chemical_2007}
Tornatore, L., Borgani, S., Dolag, K., \& Matteucci, F. 2007, MNRAS, 382, 1050

\bibitem[{Valenzuela \& Remus(2022)}]{valenzuela_stream_2022}
Valenzuela, L.~M. \& Remus, R.-S. 2022

\bibitem[{Villaescusa-Navarro {et~al.}(2020)Villaescusa-Navarro, Hahn, Massara,
  Banerjee, Delgado, Ramanah, Charnock, Giusarma, Li, Allys, Brochard,
  Uhlemann, Chiang, He, Pisani, Obuljen, Feng, Castorina, Contardo, Kreisch,
  Nicola, Alsing, Scoccimarro, Verde, Viel, Ho, Mallat, Wandelt, \&
  Spergel}]{villaescusa-navarro_quijote_2020}
Villaescusa-Navarro, F., Hahn, C., Massara, E., {et~al.} 2020, ApJS, 250, 2

\bibitem[{Vojtekova {et~al.}(2021)Vojtekova, Lieu, Valtchanov, Altieri, Old,
  Chen, \& Hroch}]{vojtekova_learning_2021}
Vojtekova, A., Lieu, M., Valtchanov, I., {et~al.} 2021, MNRAS, 503, 3204

\bibitem[{Zhang {et~al.}(2019)Zhang, Yanny, Palmese, Gruen, To, Rykoff, Leung,
  Collins, Hilton, Abbott, Annis, Avila, Bertin, Brooks, Burke, Carnero~Rosell,
  Carrasco~Kind, Carretero, Cunha, D'Andrea, da~Costa, De~Vicente, Desai,
  Diehl, Dietrich, Doel, Drlica-Wagner, Eifler, Evrard, Flaugher, Fosalba,
  Frieman, García-Bellido, Gaztanaga, Gerdes, Gruendl, Gschwend, Gutierrez,
  Hartley, Hollowood, Honscheid, Hoyle, James, Jeltema, Kuehn, Kuropatkin, Li,
  Lima, Maia, March, Marshall, Melchior, Menanteau, Miller, Miquel, Mohr,
  Ogando, Plazas, Romer, Sanchez, Scarpine, Schubnell, Serrano, Sevilla-Noarbe,
  Smith, Soares-Santos, Sobreira, Suchyta, Swanson, Tarle, Thomas, Wester, \&
  {DES Collaboration}}]{zhang_dark_2019}
Zhang, Y., Yanny, B., Palmese, A., {et~al.} 2019, ApJ, 874, 165

\bibitem[{Zibetti {et~al.}(2005)Zibetti, White, Schneider, \&
  Brinkmann}]{zibetti_intergalactic_2005}
Zibetti, S., White, S. D.~M., Schneider, D.~P., \& Brinkmann, J. 2005, MNRAS,
  358, 949

\end{thebibliography}

\end{document}